\documentclass[twoside,reqno]{qts-proc}
\usepackage{epsfig,cite}
\usepackage{amssymb,amsmath}
\usepackage{times}
\begin{document}
\sloppy \raggedbottom \setcounter{page}{1}

\newpage
\setcounter{figure}{0} \setcounter{equation}{0}
\setcounter{footnote}{0} \setcounter{table}{0}
\setcounter{section}{0}


\def\bec{\begin{center}}
\def\ec{\end{center}}
\def\a{\alpha} \def\ad{\dot{\a}} \def\ua{{\underline \a}}\def\hA{{\widehat A}}
\def\unA{\underline A}\def\unB{\underline B}\def\dB{\dot B}
\def\b{\beta}  \def\bd{\dot{\b}} \def\ub{{\underline \b}}
\def\c{\gamma} \def\cd{\dot{\c}}
\def\C{\Gamma}
\def\d{\delta} \def\dd{\dot{\d}}
\def\D{\Delta}
\def\e{\epsilon} \def\he{\hat{\epsilon}}
\def\ve{\varepsilon}
\def\f{\phi}\def\tf{\tilde{\phi}}
\def\F{\Phi}\def\tF{\tilde{\Phi}}
\def\vf{\varphi}
\def\k{\kappa}
\def\l{\lambda}\def\tl{\tilde{\lambda}}
\def\L{\Lambda}
\def\m{\mu}
\def\n{\nu}
\def\r{\rho}
\def\s{\sigma}
\def\S{\Sigma}
\def\t{\tau}
\def\th{\theta} \def\tb{\bar\theta}
\def\Th{\Theta}
\def\x{\xi}
\def\y{\eta}
\def\z{\zeta}
\def\O{\Omega}
\def\o{\omega}
\def\pb{{\bar\pi}}
\def\eb{{\bar\varepsilon}}
\def\lb{{\bar\lambda}}
\def\cB{{\cal B}}
\def\cG{{\cal G}}

\def\cL{{\cal L}}
\def\cD{{\cal D}}
\def\cF{{\cal F}}
\def\cO{{\cal O}}
\def\cA{{\cal A}}
\def\cM{{\cal M}}
\def\cN{{\cal N}}
\def\cR{{\cal R}}
\def\cS{{\cal S}}
\def\cP{{\cal G}}
\def\cV{{\cal V}}
\def\cH{{\cal H}}
\def\cI{{\cal I}}
\def\vk{{\vec k}}
\def\vx{{\vec x}}
\def\vz{{\vec z}}
\def\vw{{\vec w}}
\def\cA{{\cal A}}
\def\hcA{\hat\cA}
\def\hF{\hat{\F}}
\def\hcF{\hat{\cal F}}
\def\hd{\hat{d}}
\def\hP{\hat{\Psi}}
\def\bp{{\bar \pi}}
\def\yb{{\bar y}}
\def\zb{{\bar z}}
\def\vb{{\bar v}}
\def\ub{{\bar u}}
\def\smpl{{\tiny +}}
\def\smm{{\tiny -}}
\def\cO{{\cal O}}
\def\cit#1{\centerline{\it #1}}
\def\ra{\rightarrow}
\def\lra{\leftrightarrow}
\def\q{\quad}
\def\qq{\quad\quad}
\def\qqq{\quad\quad\quad}
\def\del{\partial}
\def\na{\nabla}
\let\la=\label
\let\bm=\bibitem
\def\nn{\nonumber}
\newcommand{\eq}[1]{(\ref{#1})}
\newcommand{\ns}[1]{{\normalsize #1}}
\def\tr{{\rm tr}}
\newcommand{\w}[1]{\\[0.#1cm]}
\def\be{\begin{equation}}
\def\ee{\end{equation}}
\def\bea{\begin{eqnarray}}
\def\eea{\end{eqnarray}}
\def\ba{\begin{array}}
\def\ea{\end{array}}
\def\se{\;\;=\;\;}
\def\seq{\;\;\equiv\;\;}
\def\dk{{d^dk\over (2\pi)^d}}
\def\intdm{\int_0^{\infty} {2\m \sinh \pi \m \over \pi^2} d\m}
\def\hf#1#2#3#4{{}_2\! F_1(#1,#2;#3;#4)}
\def\mx#1#2#3#4{\left#1\begin{array}{#2} #3 \end{array}\right#4}
\def\nab{\nabla}
\def\ft#1#2{{\textstyle{{\scriptstyle #1}
\over {\scriptstyle #2}}}} \def\fft#1#2{{#1 \over #2}}
\def\wwb#1#2#3{$\left(\fft{#1}{#3},\,\fft{#2}{#3}\,\right)$}
\def\sst#1{{\scriptscriptstyle #1}}
\def\oneone{\rlap 1\mkern4mu{\rm l}}
\def\ket#1{|#1\rangle}
\def\scs#1{\section{\sc \large #1}}
\def\scss#1{\subsection{\sc  #1}}
\def\scsss#1{\subsubsection{\sc \small #1}}

\def\ad{\dot\alpha}
\def\bd{\dot\beta}
\def\sb{\bar\sigma}




\title{On an Exact Cosmological Solution of Higher Spin Gauge Theory}

\runningheads{ E.~Sezgin and P.~Sundell}{On $SO(3,1)$ Invariant
Exact Solution of Higher Spin Gauge Theory}

\begin{start}


\author{E. Sezgin}{1},
\coauthor{P. Sundell}{2}

\address{George P. and Cynthia W. Mitchell Institute for
Fundamental Physics, \\ Texas A\&M University, College Station, TX
77843-4242, USA}{1}
\address{Department for Theoretical Physics, Uppsala
University,\\ Box 803, 751 08 Uppsala, SWEDEN}{2}


\begin{Abstract}
We review our recent exact solution to four-dimensional higher spin
gauge theory invariant under a higher spin extension of $SO(3,1)$
and we comment on its cosmological interpretation. We find an
effective Einstein-scalar field theory that admits this solution,
and we highlight the significance of the Einstein frame and what we
call higher spin frame in the cosmological interpretation of the
solution.
\end{Abstract}
\end{start}


\section{Introduction}

Four-dimensional interacting higher spin gauge theory is an
extension of ordinary gravity by an infinite tower of higher-rank
symmetric tensor gauge fields as well as particular lower spin
fields. The simplest model -- the minimal bosonic model --
consists of physical fields of rank $s=0,2,4,\dots$ with the
rank-2 field playing the role of a metric and the rank-0 field
playing the role of a particular matter sector. The field
equations can be given in a generally covariant weak-field
expansion, in which all physical fields except the metric are
treated as small fluctuations. The metric field equation contains
a negative cosmological term, and the theory admits the anti-de
Sitter spacetime as an unbroken vacuum solution, with radius set
by the fundamental length scale.

A generic feature of higher spin gauge theory is that the field
equations are strongly coupled in the sense of derivative
expansion, which means that the weak-field expansion is limited to
the perturbative study of solutions with small curvatures as well
as small scalar field fluctuations. In particular, the scalar
field potential is blurred by equally sized higher-derivative
corrections. Moreover, in the metric sector there is in general
torsion. Finally, there is in general no known consistent
truncation of the equations down to the lower-spin sector, as
lower-spin fields serve as sources for higher spin fields, and
there is no independent coupling constant that can be identified
with the higher spin fields.

Given this state of affairs, while full higher spin gauge theories
have been known in $D\leq 4$ since the early work of Vasiliev
\cite{vasiliev}, not much is known about their exact solutions
beyond the anti-de Sitter vacuum. The equations assume, however, a
remarkably simple form when written in terms of master fields --
conjectured in \cite{Johan} to be a topological open twistor
string describing the phase-space, or deformation, quantization of
the scalar $SO(3,2)$-singleton -- which are integrable in the
sense that the gauge fields, contained in a master one-form
$\widehat A_\mu$, and thus the space-time geometry can be given
\emph{algebraically} in terms of the Weyl tensors and matter
fields contained in a master zero-form $\widehat \Phi$. This
formulation becomes especially powerful when $\widehat\Phi$ is
fixed completely by symmetries, such as the cases examined in
\cite{Sezgin:2005pv} that are invariant under $3,4,6$ dimensional
groups.

Here we shall review the resulting $SO(3,1)$ invariant exact
solution found in \cite{Sezgin:2005pv} and comment on its
cosmological interpretation. In particular, we shall point out the
significance of the Einstein frame and what we call higher spin
frame in the cosmological interpretation of the solution. The later
frame naturally arises in higher spin field equations and it has
bosonic torsion, while the Einstein frame is more natural for the
cosmological interpretation. Indeed, as we shall show here, the
latter frame avoids a big crunch singularity, provided the standard
language appropriate to gravity is used, in which the notions of
horizons and singularities are based on the geodesic equation motion
of ordinary test particles. A more rigorous understanding requires,
however, a higher spin covariant counterpart, which is still not
available. Nonetheless it is our hope that our interpretation
captures some significant features of the ultimate story. We shall
also compare these results with those of \cite{Hertog:2004rz} where
an AdS cosmological solution of a consistently truncated sector of
gauged $D=4$, ${\cal N}=8$ supergravity has been examined, and a big
crunch singularity occurs.

\section{The Master Equations and the Gauge Function}

The minimal bosonic model is an extension of AdS gravity with spin
$s=0,2,4,...$ fields, each occurring once. These are exactly the
massless representations which occur in the symmetric tensor
product of two ultra-short fundamental representations of
$SO(3,2)$ known as singletons. The occurrence of a scalar field is
noteworthy and it is a universal feature of all higher spin gauge
theories.

Master fields denoted by $(A_\mu,\Phi)$ arise naturally in the
corresponding frame-like, or unfolded, formulation as follows.
Firstly, the vierbein $e_\m{}^a$ (whose relation to the Einstein
frame is discussed in Section 4), the Lorentz connection
$\o_\m{}^{ab}$, and their higher spin analogs $W_{\m,a_1\dots
a_{s-1},b_1\dots b_t}$, $0\leq t\leq s-1$, $s=4,6,\dots$, with
$W_{(a_1,a_2\dots a_s)}$ defining the physical spin-$s$ field,
make up the adjoint master one-form
 \be
 A_\m(x,y,\yb)\ =\ \frac1{2i}(e_\m{}^aP_a+\frac12
 \o_\m{}^{ab}M_{ab}+\cdots)=e_\m+\omega_\m+W_\m+K_\m\ ,
 \label{Amu}
 \ee
where $K_\m$ is a field re-definition required for manifest
$SO(3,1)$ invariance, to be described below, and $(M_{ab}, P_a)$
are the $SO(3,2)$ generators which can be realized in terms of
$SL(2,C)$-doublet oscillators $y_\a$ and $\bar
y_\a=(y_\a)^\dagger$ as
 \be
 M_{ab}\ =\ -\frac18 \left[~ (\s_{ab})^{\a\b}y_\a y_\b+
 (\bar\s_{ab})^{\ad\bd}\yb_{\ad}\yb_{\bd}~\right]\ ,\qquad P_{a}\ =\
 \frac14 (\s_a)^{\a\bd}y_\a\yb_{\bd}\ .\label{mab}
 \ee
Second, the scalar field $\phi$, the spin-2 Weyl tensor
$C_{ab,cd}$ (which we take to be symmetric in its pairs of
indices), its higher spin analogs and all possible derivatives of
these fields fit into a twisted-adjoint master zero-form
 \bea
 \Phi(x,y,\yb)&=& \phi+iP^a\nabla_a\phi +i^2P^aP^b\nabla_a\nabla_b\phi+
 \cdots\nn\\ && +~
 M^{ab}M^{cd}\left(C_{ac,bd}
 +\cdots\right)\nn\\&&+~\mbox{spin $s=4,6,...$ sectors}\ ,
 \label{Phi}
 \eea
where combinatorial coefficients are suppressed. The master fields
$A_\mu$ and $\Phi$ are extended -- or deformation quantized --
into full master fields $\widehat A$ and $\widehat \Phi$ obeying
the constraints
 \bea
 \widehat{F}&\equiv & \widehat
 d\widehat{A}+\widehat{A}\star\widehat{A}\ =\ \frac{i}4\widehat
 \Phi\star\left(b
 dz^\a dz_\a e^{iy^\a z_\a} + \bar b d\zb^{\ad} d\zb_{\ad}~
 e^{-i\yb^{\ad}\zb_{\ad}}\right)\ ,
 \nn\\[5pt]
 \widehat{D}\widehat{\Phi}&\equiv&\widehat d\widehat{\Phi}+\widehat
 A\star\widehat \Phi-\widehat \Phi\star \pi(\widehat A)= 0\
,\la{c2}\eea
with $\widehat d=d+\widehat d'$ where $d=dx^\mu\partial_\mu$ and
$\widehat d'= (dz^\a\partial_\a+{\rm h.c.})$ are exterior
derivatives on a spacetime ${\cal M}$ and a non-commutative
twistor space ${\cal Z}$, respectively. The parameter $b=1$ in
Type A model, in which the scalar $\phi$ is even under parity, and
$b=i$ in the Type B model, in which $\phi$ is odd under parity.
The extended master fields are maps from ${\cal M}\times {\cal Z}$
to the space of functions on ${\cal Z}$, \emph{viz.}
 \bea
 \widehat{A}&=&  dx^\mu \widehat{A}_\mu(x,z,\bar z;y,\bar y) +
 dz^\a \widehat{A}_\a(x,z,\bar z;y,\bar y)+
 d\zb^{\ad}\widehat{A}_{\ad}(x,z,\bar z;y,\bar y)\ ,\nn\w2
 \widehat{\Phi}&=&\widehat\Phi(x,z,\bar z;y,\bar y)\ ,
 \qquad A_\mu = \widehat A_\m|_{Z=0}\ ,\qquad \Phi = \widehat \Phi|_{Z=0}\
.\label{ic}\eea
where $(x^\m,z^\a,\bar z^{\ad};y^\a,\bar y^{\ad})$ coordinatize
${\cal M}\times {\cal Z}\times {\cal Z}$ and the associative
$\star$-product is defined by
 \bea
 &&\widehat f(y,\bar y;z,\bar z)~\star~ \widehat g(y,\bar y;z,\bar z)
 = \int \frac{d^2\xi d^2\eta d^2\bar\xi
 d^2\bar\eta}{(2\pi)^4} e^{i\eta^\a\xi_\a+ i\bar\eta^{\dot\a}\bar\x_{\dot\a}} \times
 \label{star}\\[5pt]
 &&\times~\widehat f(y+\xi,\bar y+\bar
 \xi;z+\xi,\bar z-\bar \xi)~\widehat g(y+\eta,\bar y+\bar
 \eta;z-\eta,\bar z+\bar \eta)\ .\nn
 \eea
The minimal master fields satisfy the additional discrete symmetry
conditions
\be {\tau}(\hA)= -\hA\ ,\quad \hA^\dagger =-\hA\ ,\quad\quad
{\tau}(\widehat\Phi)={\bar\pi}(\hF)\ ,\quad (b\hF)^{\dagger}=\pm b
\pi(\hF)\ ,\la{hf2}\ee
where $\tau(\widehat f(y,\yb,z,\zb))=\widehat f(i y,i\yb,-i z,-i
\zb)$ and $\pi(\widehat f)=\widehat f (-y,\yb;-z,\zb)$. The sign
in the equation involving $b$ corresponds to $\phi(x)=\widehat
\Phi|_{Y=Z=0}$ transforming under parity (acting in tangent space)
into $\pm \phi(x)$ with $+$ in Type A model and $-$ in Type B
model. By convention, we take $b=1$ and $b=i$ in the Type A and B
models, respectively, so that
$\widehat\Phi^\dagger=\pi(\widehat\Phi)$ and $\phi^\dagger=\phi$.

The gauge transformation are given by
 \be
 \delta_{\widehat\e}\widehat A=\widehat D\widehat\e\ ,\qquad
 \delta_{\widehat\e}\widehat\Phi=-\widehat\e\star\widehat\Phi+
 \widehat\Phi\star\pi(\widehat\e)\ .
 \ee
A close examination of the the Lorentz transformations of the full
master fields \cite{Vas:star,Us:analysis} given by
 \bea
 \widehat \epsilon_L &=& \frac1{4i}\Lambda^{\a\b}(x)\widehat M_{\a\b}-{\rm h.c.}\
 ,\\
 \widehat{M}_{\a\b} &=&  y_\a y_\b - z_\a z_\b +\frac12 \,\{
 \widehat S_\a,\widehat S_\b\}_*\ ,\quad \widehat{S}_\a\ =\
 z_\a-2i\widehat{A}_\a\ ,
 \label{Mhat}
 \eea
shows that $e_\mu{}^a$ and $W_\mu$, defined in \eq{Amu}, transform
canonically under the Lorentz transformation provided
 \be
 K_\m \ =\  {1\over4i}\o_\m{}^{\a\b}\widehat S_\a\star\widehat
 S_\b|_{Z=0}-{\rm h.c.}\ =\ i\o_\m{}^{\a\b}
 \left.(\widehat{A}_\a\star\widehat{A}_\b)\right|_{Z=0}-\mbox{h.c.}\ ,
 \la{k}
 \ee
where the gauge condition $\widehat A_\a|_{\widehat\Phi=0}=0$ has
been assumed. Thus, locally, a space-time field configuration
$\phi(x)$, $g_{\m\n}(x)$ and $\phi_{\m_1\dots \m_s}(x)$
($s=4,6,\dots$) can be unfolded and packed into a twisted-adjoint
initial condition $\Phi(x;y,\yb)|_{x=0}$, which is deformed into
 \be
 \widehat\Phi'(z,\zb;y,\yb)=\widehat \Phi|_{x=0}\ ,\qquad \widehat
 A'_\a(z,\zb;y,\yb)=\widehat A_\a(x,z,\zb;y,\yb)|_{x=0}\ .
 \ee
This can be made precise by solving the constraints $\widehat
F_{\m\n}=0$, $\widehat F_{\m\a}=0$ and $\widehat D_\m
\widehat\Phi$ using a gauge function $\widehat L=\widehat
L(x,z,\zb;y,\yb)$,
 \be
\widehat A_\mu\ =\ \widehat L^{-1}\star \partial_\mu \widehat L\
,\qquad \widehat A_\a\ =\ \widehat L^{-1}\star (\widehat A'_\a+
\partial_\a) \widehat L\ ,\qquad \widehat \Phi\ =\ {\widehat L}^{-1}\star
\widehat\Phi'\star \pi(\widehat L)\ ,\label{Leq}
 \ee
and determine the remaining $Z$-dependence from
\bea \widehat
F'_{\a\b}&\equiv&2\partial^{\phantom{\prime}}_{[\a}\widehat
A'_{\b]}+[\widehat A'_\a,\widehat A'_\b]_\star\ =\ \ -\frac{ib}2
\e_{\a\b}\widehat\Phi'\star \kappa\ ,\label{z1}\\[5pt]\widehat F'_{\a\bd}&\equiv&
\partial^{\phantom{\prime}}_\a\widehat A'_{\bd}-\partial_{\bd}\widehat
A'_{\a}+[\widehat A'_\a,\widehat A'_{\bd}]_{\star}\ =\ 0\ ,\label{z2}\\[5pt]
\widehat D'_\a\widehat \Phi'&\equiv&
\partial^{\phantom{\prime}}_\a\widehat \Phi'+\widehat
A'_\a\star\widehat\Phi'+\widehat \Phi'\star\pi(\widehat A'_\a)\ =\
0\ ,\label{z3}\eea
given $\widehat\Phi'|_{Z=0}\equiv C'(y,\bar y)$ and fixing the
gauges $\widehat A'_\a|_{C'=0}=0$ and $\widehat
L|_{C'=0}=L(x;y,\yb)$, in turn implying $\partial_\a\widehat L=0$,
that is, $\widehat L=L(x;y,\yb)$.

In case $\widehat\Phi'$ is invariant under a symmetry group $G_r$
with full parameters $\widehat\e'$, and assuming that $\widehat
\Phi'$ and $\widehat \e'$ have well-defined perturbative
expansions in $C'$ of the form $\widehat
\Phi'=C'+\widehat\Phi'_{(2)}+\cdots$ and
$\widehat\e'=\e'+\widehat\e'_{(1)}+\cdots$, where $\e'=\e'(y,\bar
y)$ is an adjoint representation of $G_r$, then it follows that
$C'$ must obey $\e'\star C'-C'\star\e'=0$. For $G_6$, the latter
condition admits two-parameter solution spaces except at the
special point \cite{Sezgin:2005pv}
\be G_6\ =\ SO(3,1)\ :\quad \e'\ =\ \frac1{4i}
\L^{\a\b}M_{\a\b}-{\rm h.c.}\ ,\quad C'\ =\ {\nu\over b}\
,\label{linso31}\ee
where $\nu/b$ is a constant real deformation parameter (which
requires $\nu$ to be real and purely imaginary in the Type A and
Type B models, respectively). Next we turn to the promotion of
this linearized solution into the exact solution given in
\cite{Sezgin:2005pv} -- which is presently the only known exact
solution to Vasiliev's four-dimensional higher spin gauge theory
other than AdS spacetime.


\section{The $SO(3,1)$ Invariant Exact Solution}

To describe the $SO(3,1)$ invariant solution in spacetime it is
convenient to use the stereographic coordinate on $AdS_4$ with
inverse radius $\l$, \emph{viz.}
 \bea
 && e_{(0)}{}^{\a\ad}= -{\l(\s^a)^{\a\ad}dx_a\over h^2}\ ,\qquad
 \o_{(0)}{}^{\a\b}\ =\ - {\l^2(\s^{ab})^{\a\b} dx_a x_b\over h^2}\
 ,\nn\w2
 && h=\sqrt{1-\lambda^2x^2}\ ,\quad x^2\ =\ x^a x^b\eta_{ab}\ ,
 \label{adseo}
 \eea
in turn corresponding via $L^{-1}\star dL={1\over
4i}(\o_{(0)}^{\a\b}y_\a
y_\b+\bar\o_{(0)}^{\ad\bd}\yb_{ad}\yb_{\bd}+2e_{(0)}^{\a\ad}y_\a
\yb_{\ad})$ to the gauge function \cite{Bolotin:1999fa}
 \be
 L(x;y,\yb)={2h\over 1+h} \exp\left[{i\lambda
 x^a(\s_a)^{\a\ad} y_\a \bar y_{\dot\a}\over 1+h}\right] \
 \label{wL2}\ ,
 \ee
with $L^{-1}(x;y\yb)=L(-x;y\yb)$. The full $SO(3,1)$ invariance
condition \eq{linso31} then becomes
\be [\widehat M'_{\a\b},\widehat \Phi']_\pi\ =\ 0\ ,\qquad
\widehat D'_\a\widehat M'_{\b\c}\ =\ 0\ ,\label{inv2}\ee
where $\widehat M'_{\a\b}$ are defined by \eq{Mhat} with internal
connection given by $\widehat A'_\a$. Using also the
$\tau$-invariance condition on $\widehat A'_\a$, it follows that
\be \widehat \Phi'\ =\ f(u,\bar u)\ ,\quad \widehat S'_\alpha\ =\
z_\a~S(u,\bar u)\ ,\qquad u\ =\ y^\a z_\a\ ,\qquad \bar u\ =\
u^\dagger\ =\ \bar y^{\dot \a} \bar z_{\dot \a}\ ,\ee
where $f$ is a real function. The internal constraints $\widehat
F'_{\a\dot\a}=0$ and $\widehat D'_\alpha \widehat \Phi'=0$ are
then solved by
\be \widehat\Phi'(u,\bar u)\ =\ {\nu\over b}\ ,\quad S(u,\bar u)\
=\ S(u)\ ,\label{sol1}\ee
where $\nu/b$ is the constant introduced in \eq{linso31}. The
remaining constraint \eq{z1} then takes the form $[\widehat
S^{\prime\alpha},\widehat S'_\a]_\star\ =\ 4i(1- \nu e^{iu})$. To
solve this equation, following \cite{Prokushkin:1998bq}, we use
the integral representation
\be S(u)\ =\ 1+ \int_{-1}^1 dt~ q(t) ~e^{\frac{i}2(1+t)u}\ ,\ee
where $t\in [-1,1]$, as can be seen from perturbation theory. The
equation for $S$ then takes the form of an integral equation that
can be solved by means of algebraic techniques invented in
\cite{Prokushkin:1998bq} (see also \cite{Sezgin:2005pv} for a
slight refinement of the basis of functions on $[-1,1]$). The
result reads
 \bea
 q(t) &=& -\frac{\nu}4
 \left(F\left({\nu\over 2}\log \frac 1{t^2}\right)
 +t \,F\left(-{\nu\over 2}\log \frac 1{t^2}\right)\right)\ ,
 \nn\\
 \quad F(\zeta)& \equiv & ~{}_1\! F_1\left[\frac12;2;\zeta\right]\
 =\ 1+{\zeta\over 4}+{\zeta^2\over 16}+
 +\cdots \ ,
 \eea
and the internal primed solution is thus given by
 \bea
 \widehat \Phi'&=&{\nu\over b}\ ,
 \label{intconn}\\
 \widehat A'_\a\ &=&\  \frac{i\nu}8 z_\a \int_{-1}^1 dt
 e^{\frac{i}2(1+t)u}\left[F\left({\nu\over 2}\log \frac 1{t^2}\right)+ t
 F\left(-{\nu\over 2}\log \frac 1{t^2}\right)\right]\ .\nn
 \eea
Expanding $\exp ({itu\over 2})$ yields integrals that converge at
$t=0$ and $t=\pm 1$, and $\widehat A_\a$ is a power-series
expansion in $u$ with coefficients that are functions of $\nu$
that are analytic at $\nu=0$ and with different analytic structure
on the real and imaginary axis.

The physical scalar field and the (auxiliary) Weyl tensors are
obtained by unpacking $\Phi=\widehat \Phi|_{Z=0}$ according to
\eq{Phi}. From
\be \widehat \Phi\ =\ L^{-1}\star \widehat \Phi'\star\pi(L)\ \
={\nu\over b} L^{-1}\star L^{-1}\ =\ {\nu\over b}
(1-\lambda^2x^2)\exp\left[-i\l x^{\a\ad}y_\a \yb_{\ad}\right]\
,\label{wphi}\ee
it follows that
\be \phi(x)\ =\ {\nu\over b} h^2\ =\ {\nu\over b}(1-\lambda^2x^2)\
, \label{scalarsol}\ee
while all Weyl tensors vanish. The above expressions are valid for
$l^2x^2<1$. The gauge fields are obtained by unpacking
$A_\mu=L^{-1}\star\partial_\mu L=e^{(0)}_\mu+\omega^{(0)}_\mu$ using
the decomposition \eq{Amu}, \emph{i.e.} $A_\m\equiv
e_\m+\o_\m+W_\m+K_\m$ with $K_\mu=i (\o_\m{}^{\a\b} L^{-1}\star
\widehat A'_\a\star \widehat A'_\b\star L+{\rm h.c.})$ given by
 \be
 K_\mu={Q \over 4i} \o_\m{}^{\a\b}\left[(1+a^2)^2y_\a
 y_\b+4(1+a^2)a_\a{}^{\ad}y_\b\yb_{\ad}+4a_\a{}^{\ad}
 a_\b{}^{\bd}\yb_{\ad}\yb_{\bd}\right]-{\rm h.c.}\ ,
 \label{Kmu}
 \ee
where $a_{\a\ad}=(1+h)^{-1}\l x_{\a\ad}$ and
\be Q\ =\ -\frac14 (1-a^2)^2\int_{-1}^1 dt\int_{-1}^1 dt'~ {
q(t)q(t')(1+t)(1+t')\over (1-tt' a^2)^4}\ .\label{Q}\ee
Decomposing $Q=Q_+(a^2)+Q_-(a^2)$, $Q_\pm(-a^2)=\pm Q_\pm(a^2)$,
one finds
\bea Q_+&=&-{(1-a^2)^2\over 4}\sum_{p=0}^\infty {-4\choose
2p}a^{4p}\left(\sqrt{1-{\nu\over 2p+1}}-\sqrt{1+{\nu\over
2p+3}}\right)^2\label{Qpl}\\[8pt]
Q_-&=&{(1-a^2)^2\over 4}\sum_{p=0}^\infty {-4\choose
2p+1}a^{4p+2}\left(\sqrt{1-{\nu\over 2p+3}}-\sqrt{1+{\nu\over
2p+3}}\right)^2\ ,\nn\eea
which have branch cuts along the real axis for ${\rm Re}~\nu\leq
-3$ and ${\rm Re}~\nu\geq 1$. From \eq{Amu} and \eq{Kmu} it
follows that all higher spin gauge fields vanish,
\be W_\m{}^{a_1\cdots a_{s-1}}\ =\ 0\ ,\quad s=4,6,...,\infty \ee
while the vierbein and Lorentz connection are given by
 \bea
 && \o^{\a\b}= f \o_{(0)}^{\a\b}\ ,\qquad
 e^a\ =\ f_1 dx^a + \l^2 f_2 dx^b  x_b x^a\ ,
 \la{s1}\w2
  && f = {1+(1-a^2)^2\bar Q\phantom{{\hat f}\over {\hat f}}\over
 |1+(1+a^2)^2Q|^2-16a^4|Q|^2\phantom{{\hat f}\over a}}\ ,
 \label{s2}\w2
 && f_1+\lambda^2 x^2 f_2= {2\over h^2}\ ,
 \qquad f_2={8(Q f+\bar Q\bar f)\over h^2(1+h)^2}\ ,
 \label{s3}
 \eea
where $a^2 =(1-h)/(1+h)$ and we recall that $h = \sqrt{1-\lambda^2
x^2}$ so that $a^2\in [-1,1]$ as $x^a$ varies over the
stereographic coordinate chart. For the Type A model, the function
$Q$ is real, and we have the simplifications
 \bea
 && f_1 = {2f\over h^2}\left[ 1+(1-a^2)^2Q\right]\ ,
 \quad f_2 = {16Qf\over h^2(1+h)^2}\ ,\nonumber\\
 && f=\left[1+(1+6a^2+a^4)Q\right]^{-1}\ .
 \eea
which are valid also to order $\nu^2$ in the Type B model.
Expanding $Q(a^2,\nu)=\sum_{n=2}^\infty \nu^n Q_n(a^2)$, the
coefficients $Q_n(a^2)$ with $n\geq 4$ are bounded while
$Q_{2,3}(a^2)$ diverge logarithmically at $a^2=-1$, as can be seen
from
 \bea Q_2& =& {(1-a^2)^2\over 48 a^4}\left[1-{2a^2\over (1-a^2)^2}
 +{(1-a^2)^2\over 2a^2}\log
{1-a^2\over 1+a^2} \right]\ ,\label{q2}\\
 Q_3 &=& {(1-a^2)^2\over 96 a^6}\left[a^2+(1-a^4)Li^{(-)}_2(a^2)+(1+a^4)
 \log{1-a^2\over 1+a^2}\right]\ ,\quad\eea
where $Li^{(-)}_2(z)=\frac12
\left(Li_2(z)-Li_2(-z)\right)=\sum_{k=0}^\infty {z^{2k+1}\over
(2k+1)^2}$. At $a^2=1$, the double integral in \eq{Q} diverges at
$t=t'=\pm 1$ while the pre-factor vanishes, producing the finite
residue
 \be
 \lim_{a^2\ra 1} Q\ =\ \lim_{a^2\ra 1} Q_2\ =\ -{\nu^2\over 24} \
 .
 \ee
Thus, for $\nu\ll1$, we can approximate $Q\simeq Q_2$ for $-1\leq
a^2\leq 1$.

To obtain a globally well-defined solution, one introduces a second
gauge function $\tilde L=L(\tilde x;y,\yb)$ where $\l^2 \tilde
x^2<1$, and $\tilde x^a=x^a/(\l^2 x^2)$ for $\l^2 x^2<0$. The two
local representatives have the same functional form, with $x^a$
replaced by $\tilde x^a$, and they are related in the overlap region
by a simultaneous reparametrization \emph{and} locally defined gauge
transformation with gauge function $\tilde L^{-1}\star L$. This
implies a $Z_2$-duality transformation, which acts on the scalar
field as \cite{Sezgin:2005pv}
\bea \widetilde\phi(\tilde x)&=&{\nu\phi(x)\over \phi(x)-\nu}\
,\qquad \lambda^2 x^2=(\lambda^2\widetilde x^2)^{-1}<0\
.\label{duality}\eea

\section{Holographic and Cosmological Interpretation}\label{sec:5}

The solution consists of a scalar field profile on a Weyl-flat
metric, which can be written as (from here on we set $\lambda=1$
for notational simplicity)
\bea ds^2&=&{4\Omega^2 (d(g_1 x))^2\over (1-g_1^2 x^2)^2}\ ,\\
\Omega&=&{(1-g_1^2 x^2)f_1\over 2g_1}\ ,\qquad g_1\ =\ {\rm
exp}~\left(\frac12 \int_1^{x^2} {f_2(t)~dt \over f_1(t)}\right)\
.\eea
The spacetime decomposes into three-dimensional $SO(3,1)$ orbits
describing local foliations of $AdS_4$ with $dS_3$ and $H_3$
spaces in the regions $x^2>0$ and $x^2<0$, respectively. In the
coordinates
 \bea x^2>0&:&
 x^0\ =\ \sinh\tau\tan {\psi\over 2}\ ,\quad
 x^i\ =\ n^i \cosh\tau\tan {\psi\over 2}\ ,\\
 x^2<0&:&
 x^0\ =\ \cosh\psi\tan {\tau\over 2}\ ,\quad
 x^i\ =\ n^i \sinh\psi\tan {\tau\over 2}\ ,
 \eea
with $n^i n^i=1$, our solution takes the form
 \bea
 x^2>0&:& ds^2\ =\ d\psi^2 +\eta^2 \sinh^2\psi\left(-d\tau^2+\cosh^2\tau~ d\Omega_2\right)\
 ,\\&& \phi\ =\ {\nu\over b} \,{\rm sech}^2{\psi\over 2}\ ,\\
 x^2<0&:& ds^2\ =\ -d\tau^2+ \eta^2 \sin^2\tau \left(d\psi^2
 +\sinh^2\psi~ d\Omega_2\right)\ ,\\&& \phi\ =\ {\nu\over b}\, {\rm sec}^2{\tau\over 2}\
 ,\eea
where
 \be
 \eta\ =\ {f_1h^2\over 2}\ ,\qquad a^2\ =\ \left\{\ba{ll}
 \tanh^2{\psi\over 4}& x^2>0\\[3pt] -\tan^2{\tau\over
 4}&x^2<0\ea\right.\ .
 \label{eta}
 \ee
In the Type A model, and to order $\nu^2$ in the Type B model, we
have
\be \eta\ =\ {1+(1-a^2)^2Q\over
 1+\left(1+6a^2+a^4\right)Q}\ .\label{48}\ee
The solution has non-trivial torsion
\be T^a\ \equiv \ de^a+\omega^a{}_b\wedge e^b\ =\ -e^a\wedge
d\log\eta\ ,\ee
that can be removed by going to an Einstein frame via a Weyl
rescaling
\be \tilde e^a\ =\ \eta^{-1}e^a\ ,\quad \tilde
\omega_{ab}=\omega_{ab}\ .\label{rescale}\ee
The resulting torsion free Einstein metric reads
\bea d\tilde s^2&=&{4\widetilde \O^2 d\tilde x^2\over (1-\tilde
x^2)^2}\ ,\quad \widetilde\Omega\ =\ {\Omega\over \eta}\ ,\eea
or in terms of foliations,
\bea
 x^2>0&:& d\tilde s^2\ =\ d\tilde\psi^2 +\sinh^2\psi
 \left(-d\tau^2+\cosh^2\tau~ d\Omega_2\right)\
 ,\\&& d\tilde\psi\ =\ {d\psi\over\eta}\ ,\\
 x^2<0&:& ds^2\ =\ -d\tilde\tau^2+ \sin^2\tau \left(d\psi^2
 +\sinh^2\psi~ d\Omega_2\right)\ ,\\&& d\tilde\tau\ =\ {d\tau\over\eta}\
 ,\eea
We propose that the Weyl rescaling \eq{rescale} can be generalized
to a background covariant transformation taking higher spin frame
\eq{Amu}, which in general has torsion $T^a$ depending on
$\widehat \Phi$, to an Einstein frame $\widetilde e^a$ in which
$\widetilde T^a=0$. Although the transformation may be complicated
in general, it should reduce to the above Weyl rescaling on the
$SO(3,1)$ invariant solution. For consistency, it must therefore
be possible to write $\eta=\eta(x^2)$ as a local background
covariant functional independent of $\nu$. Indeed, as found in
\cite{Sezgin:2005pv} there exist zero-forms ${\cal
C}^{-}_{(2n)}={\cal C}^{-}_{(2n)}[\widehat\Phi]$ that reduce on
the $SO(3,1)$ invariant solution to $\nu^{2n}$, which can then be
used to define the Weyl rescaling covariantly by taking
$\eta=\eta((1-\phi)/{\cal C})$ by choosing, for example, ${\cal
C}[\widehat \Phi]=\sqrt{{\cal C}^-_{(2)}}$.

In the asymptotic region $x^2\rightarrow 1$, the scale factor
$\widetilde\Omega\rightarrow 1$ and the scalar field
\be \phi\ =\ {\nu\over b}(\xi-\frac12 \xi^2+\cdots)\ ,\ee
where the radial coordinate $\xi$ is defined by $\tanh^2
(\psi/2)=e^{-\xi}$, and the unperturbed $AdS_4$ metric reads
$ds_{(0)}^2=(dr^2+ds^2_{dS_3})/\sinh^2(\xi/2)$. In global
coordinates, $ds_{(0)}^2=-(1+r^2)dt^2+{dr^2\over
1+r^2}+r^2d\Omega^2$, one instead finds
\be \phi\ =\ {2\nu\over b}\left( {1\over r\sin t}-{1\over
r^2\sin^2 t}+\cdots\right)\ .\ee
In general, if $\phi=\a z+\beta z^2+\cdots$ and
$ds^2_{(0)}=(dz^2+d\sigma^2)/(\l(z))^2$ where $\l$ has a simple zero
at $z=0$, then the relation $\b=\b(\a)$ describes a deformation of
the holographically dual field theory, which has been conjectured to
be the $O(N)$ model and the Gross-Neveu model in the cases of the
Type A and Type B models, respectively
\cite{holography,PK,PL,cubic}. In our case, we find
\be \beta\ =\ -k\a^2\ ,\quad k\ =\ {b\over 2\nu}\ ,\label{bc}\ee
corresponding to a marginal triple-trace deformation of the
ultraviolet fixed points of the $O(N)$ and GN models built from the
scalar Konishi operator along the lines discussed in
\cite{Hertog:2004rz}. Interestingly, by considering a quantum
mechanical approximation, the deformation was found in
\cite{Hertog:2004rz} to generate a bounce in the expectation value
of the Konishi operator. Indeed, this is in qualitative agreement
with the scalar field profile traced out by the bulk scalars
$\phi(x)$ for $\l^2 x^2>-1$ and $\widetilde\phi(\tilde x)$ for
$\l^2\tilde x^2>-1$. In \cite{Hertog:2004rz} this deformation was
considered as a simplified model, obtained essentially by neglecting
the non-abelian structure on the D2 brane, meant to capture
qualitatively the behavior of an analogous marginal triple-trace
deformation of the three-dimensional CFT on coinciding membranes
forming the holographic dual of a $SO(3,1)$-invariant instanton of
gauged ${\cal N}=8$ supergravity. Here we instead consider it as the
actual holographic dual of our solution to the higher-spin gauge
theory.

The minimal bosonic models we have studied here are consistent
truncations of the higher spin gauge theory based on
$shs(8|4)\supset osp(8|4)$ \cite{hs1}, which contains, respectively,
$35_++35_-$ scalars and pseudo-scalars in the supergravity multiplet
and $1+1$ scalar and pseudo-scalar in an $s_{\rm max}=4$ multiplet,
which we refer to as the Konishi multiplet. While our solutions in
the Type A and Type B models utilize the Konishi scalar and
pseudo-scalar, respectively \cite{Engquist:2002vr}, the solution of
\cite{Hertog:2004rz} activates instead one of the scalars residing
in the supergravity multiplet. Therefore a meaningful comparison of
the solutions requires two steps. First, the construction of a new
solution in which one of the supergravity scalars in the higher spin
gauge theory is activated. Second, the higher spin symmetries must
be spontaneously broken down to standard diffeomorphisms.

The breaking of higher spin symmetries requires Goldstone modes
which can either be fundamental
\cite{sundborg,sundell2,Beisert:2004di} or composite
\cite{polyakovklebanov,porrati}. In the former scenario, the
symmetries are broken classically by the stringy dilaton and the
Goldstone modes are massive multi-singleton states. This corresponds
to a non-abelian D2-brane deformation of the holographic dual,
whereby the higher spin multiplets are separated from the
supergravity multiplet by a large mass-gap. In the latter scenario,
on the other hand, the non-abelian structures are not activated, and
the virtual processes are instead implemented on the field theory
side in the form of double-trace-like ``sewing operations''
\cite{sundell2,polyakovklebanov}. Correspondingly, radiative
corrections in the bulk induce small mass gaps provided the Konishi
scalars are subjected to suitable boundary conditions
\cite{porrati}. It should, of course, also be possible to quantize
the theory while preserving all symmetries by imposing other
boundary conditions (reflecting the conformal dimensions at the free
fixed point).

This suggests that M theory on $AdS_4\times S^7$ with $N$ units of
seven-form flux has two phases: a supergravity phase, where all
higher spin symmetries are strongly broken, and a higher spin phase,
where all symmetries are either unbroken or weakly broken by
radiative corrections as described above. In the supergravity phase
there are two mass-scales: the Planck scale and the membrane scale,
given by powers of $N$ such that the latter is much smaller than the
former. Since $N$ is the only free parameter, one may therefore
speculate that the supergravity phase arises for energies much
smaller than the membrane scale, while the higher spin phase arises
for energies much larger than the membrane scale and much smaller
than the Planck scale, such that the membrane is effectively
tensionless while the bulk theory is nonetheless weakly coupled. The
resulting spectrum should contain massless as well as massive
fields, that we expect arise from the tensionless membrane along the
lines discussed in \cite{Johan}. We think of our solution as exact
in the classical $N\rightarrow \infty$ limit of the higher spin
phase.

With the caveats mentioned in the Introduction in mind, having to do
with the lack of an understanding of higher-spin covariant geometry,
we next proceed to examine some salient features of the standard
geometry of our solution. For $x^2\geq 0$, all the scale factors
remain finite and non-vanishing. At $x^2=0$, the scale factors
$\eta\sinh\psi$ and $\eta\sin \tau$ have $\psi$ and $\tau$
derivatives equal to $1$, respectively, which means that the DW
region is ``glued'' smoothly to the FRW regions (without deficit or
excess angle). For $\nu\ll 1$ and $1+a^2\ll 1$, we can approximate
$Q\simeq -\frac{\nu^2}{6}~\log{1\over 1+a^2}$. Thus, for $\tau\sim
\tau_{\rm crit}$, given by
 \be
 \sin{\tau_{\rm crit}}\ \simeq\  e^{-3\over 2|\nu^2|} \label{crit}
 \ee
the scale factor $\eta$ behaves as
\bea \eta&\sim&\left[{|\nu^2|\over 3}e^{3\over 2|\nu|^2}(\tau_{\rm
crit}-\tau)\right]^{\e}\ ,\quad \e\ =\ \left\{\ba{ll}+1&\mbox{A
model}\\[3pt] -1&\mbox{B model}\ea\right. \eea
Thus, in the Type A model it takes infinite proper time (measured
in Einstein frame) to reach the critical point defined in
\eq{crit}, where we note that the scalar field takes the value
 \be
 \phi_{\rm crit}\ \simeq\ {4\nu\over b}e^{3\over |\nu^2|}\ .
  \ee
On the other hand, in the Type B model it takes finite proper time
to pass this point and eventually reach $\tau=\pi$, which is the
surface where $\phi\ra+\infty$. Beyond this surface $h^2$ is
negative, as can be seen either by going to global coordinates or
taking $x^2>1$, which makes the gauge function $L$ and hence the
solution formally ill-defined (at $\tau=\pi$ the first derivatives
of the FRW scale factors $\eta\sin\tau$ and $\sin\tau$ with respect
to $\tau$ and $\tilde\tau$, respectively, are equal to $+1$, while
their higher derivatives blow up, so that the scale factors are not
real analytic at $\tau=\pi$). Thus, the $SO(3,1)$ invariant
cosmology is singularity free in the Type A model, in the sense that
it takes infinite proper time to reach the critical point, while it
hits the singularity at $\tau=\pi$ in finite proper time in the Type
B model. This phenomenon may eventually be understood starting from
the microscopic origin of the Vasiliev equations based on
topological open phase-space strings \cite{Johan}. These probe the
$SO(3,1)$ invariant phase-space geometry described by $\widehat
\Phi'$ and $\widehat A'_\a$, which appears to be singularity free in
both the Type A and Type B models.

\section{The Effective Einstein-Scalar Field Theory}

The qualitative features of the solution to the higher spin gauge
theory can be reproduced by a standard scalar-coupled gravity model.
To construct an ``effective'' action whose field equations admit the
solution presented above, we proceed as follows. We begin by
parametrizing the Lagrangian as
 \be
 e^{-1}{\cal L}= K \left( R(\o)  -\frac12 G
 \partial_\mu\phi\partial^\mu\phi -V\right)\ ,
 \ee
where $K,G,V$ are functions of $\phi$ to be determined. In this
section we set $\lambda=1$, which can be easily re-instated by
dimensional analysis, for notational simplicity. We work in first
order formalism and therefore treat the spin connection $\o$ as an
independent field. Thus, the field equations are
 \bea
 R_{\mu\nu}(e) &=&\frac12 G\partial_\mu\phi\partial_\nu\phi
 + \frac12 Vg_{\mu\nu}\ ,
 \label{fe1}
 \w2
 T_{\mu\nu}{}^a &=& e_{[\mu}^a \partial_{\nu]} \log K\ ,
 \label{fe2}
 \eea
where the torsion tensor is defined as usual by $T^a=d e^a
+\omega^a{}_b\wedge  e^b$, and $R_{\mu\nu}(e)$ is the symmetric
part of $R_{\mu,\nu}(\o)= R_{\mu\rho}{}^{ab}(\o) e^\rho_b\,e_{\nu
b}$, and as such, it is the standard Ricci tensor in terms of
torsion-free spin connection, or equivalently, the symmetric
Christoffel symbol. In obtaining \eq{fe1}, we have used \eq{fe2}
to show that $R_{[\mu,\nu]}=0$. The equation \eq{fe2} follows from
the variation of the action with resect to the spin connection. As
for the scalar field equation, it follows by taking the divergence
\eq{fe1} and using the conservation law $D^\mu(R_{\mu\nu} -\frac12
g_{\mu\nu} R)=0$.

Substituting our solution into the field equations, after
considerable algebra we find for the Type A model that
 \bea
 K &=& {4\over (f_1h^2)^{2}}\ , \w2
 G &=& 2 K \left(f_1 {\partial
 f\over\partial h^2 } +\frac12 f_2 f\right)\ ,
 \w2
 V&=& -6 K \left[ f^2 +f(1-f)h^2 \right] +\frac 12 h^4(1-h^2) G \ ,
 \eea
where it is understood that $h^2=\phi/\nu$. Next, we observe that
the Hilbert-Einstein term can be written in terms of torsion free
connection $\o_\mu{}^{ab}(e)$ by using the relation $
\omega_\mu{}^{ab}=\omega_\mu{}^{ab}(e) + e_\mu{}^{[a}\partial^{b]}
{\partial \log K\over\partial \phi}$, and subsequently we can go
over to Einstein frame by rescaling the metric as
 \be
 g_{\mu\nu}= K^{-1}~{\bar g}_{\mu\nu}\ .
 \ee
Note that, evaluated on the solution, $K^{-1}=\eta^2$, with $\eta$
given in \eq{48}. Dropping the bar for simplicity in notation, we
get the action
 \be
I= \int d^4 x~{\sqrt -g}  \left( R(\o(e))
 -\frac12 G \partial_\mu\phi\partial^\mu\phi- \frac1{4} K^{-1} V \right)\ ,
 \ee
Evidently the potential takes a highly complicated form,
reflecting the higher derivative scalar field self-couplings in
the full theory, which can be rewritten as contact terms on the
$SO(3,1)$ invariant solution. Since the potential is obtained to
accommodate our exact solution, it offers a highly limited
information regarding the structure of the full action or field
equations of the higher spin gauge theory. Nonetheless, the
potential is not fixed by picking just any forms of metric and
scalar field configurations we like (though this approach may be
of some utility in its own right for gaining insights to some
aspects of gravitational instantons, as shown, for example, in
\cite{Mukherjee:1992gs}), but rather it is a consequence of a
solution that is dictated \emph{\'a priori} by a well defined
higher spin gauge theory.

\section{Concluding remarks}

Higher spin gauge theory, which at a first superficial glance may
appear to be more complex than ordinary gravity, in fact exhibits
a remarkable simplicity in that the field equations can be solved
by means of purely algebraic methods. Thus, nontrivial exact
solutions can be given even without knowing the action nor the
equations of motion in a form in which the spacetime fields and
their couplings are explicitly displayed. One may speculate that
Vasiliev's equations are somehow exactly solvable in phase space
(see \cite{Prokushkin:1998bq} for concrete work along these lines)
so that any solution could be obtained algebraically starting from
the knowledge of the Weyl zero-form at a point in space time.
Ultimately, once the connection to ordinary gravity has been made
more explicit, one may hope that these basic properties of higher
spin gauge theory could also shed light on similar issues in
ordinary gravity.

As for the cosmological applications of our exact solution, which is
the only known time dependent solution of the higher spin gauge
theory at present, a further development of the theory is needed to
provide a geometric formulation with manifest higher spin
symmetries, and to  facilitate the description of geodesic equation
of motion and harmonic analysis. Recalling that the higher spin
gauge theory is believed to emerge from tensionless limit of strings
and branes, it is tempting to envisage a new cosmological model in
which the very early universe is described by tensionless strings
and branes, and that as the universe cools down, the higher spin
symmetries first break down by mechanism mentioned earlier to the
usual symmetries associated with spin $s\le 2$ massless fields, and
subsequently the universe evolves in the more familiar fashion that
involves inflation and other phenomena that we can describe by means
of matter coupled supergravity theories embedded in the tensionful
broken phase of string theory. Matter couplings can be described in
higher spin gauge theory even though the formalism for achieving
this has not been adequately developed so far. Concerning inflation
in the context of asymptotically AdS spacetimes, the idea that such
spacetimes may contain an inflationary de Sitter regions (see, for
example, \cite{Freivogel:2005qh} and references therein) may also be
entertained in the context of a cosmological model based on higher
spin gauge theory. In particular, it would be interesting to
determine whether the massless Konishi scalars that are present only
in the tensionless limit have a role to play in any inflation
scenario.

%


\section*{Acknowledgments}

We thank Thomas Hertog, Gary Horowitz, Jason Kumar, Hong Lu, Chris
Pope and Misha Vasiliev for useful discussions. P.~Sundell would
like to thank the George P. and Cynthia W. Mitchell Institute for
Fundamental Physics for hospitality. The work of E.S. is supported
in part by NSF grant PHY-0314712 and the work of P.S. is supported
in part by INTAS Grant 03-51-6346.


\end{document}